# Minimum and terminal velocities in projectile motion

E. N. Miranda,* S. Nikolskaya and R. Riba

Facultad de Ingeniería

Universidad de Mendoza

Arístides Villanueva 750

5500 - Mendoza, Argentina

August 2, 2004

## Abstract

The motion of a projectile with horizontal initial velocity $V_0$, moving under the action of the gravitational field and a drag force is studied analytically. As it is well known, the projectile reaches a terminal velocity $V_{term}$. There is a curious result concerning the minimum speed $V_{min}$; it turns out that the minimum velocity is lower than the terminal one if $V_0 > V_{term}$ and is lower than the initial one if $V_0 < V_{term}$. These results show that the velocity is not a monotonous function.

If the initial speed is not horizontal, there is an angle range where the velocity shows the same behavior mentioned previously. Out of that range, the velocity is a monotonous function. These results comes out from numerical simulations.

---

*Permanent addresses: CRICYT-CONICET, 5500 Mendoza, Argentina and Depto. de Fisica, Univ. Nac. de San Luis, 5700 San Luis, Argentina.

In this article we deal with a simple problem, namely to determine the velocity $V$ of a projectile launched with horizontal velocity in a medium with a drag force proportional to a power $n$ of the body speed. The motivation to deal with such a problem is pedagogical. On one hand, it fits well in an intermediate course on classical mechanics [1] or even in a general physics course with calculus [2]. On the other hand, it could be use as a physical example of the Bernoulli equation in a mathematical physics course [3] or as an exercise in numerical analysis with an interesting physical interpretation.

It will be shown that the terminal velocity $V_{term}$ is different from the minimum one $V_{min}$. Naively, one would have expected that the minimum velocity is the terminal one, but it comes out that the projectile speed diminishes until it reaches a minimum and then it starts increasing. The terminal velocity is finally reached from below, i.e. from lower velocities. This means that the velocity is not a monotonous function but it shows a minimum different from the initial and terminal velocities. This behavior is the same for any value of the power $n$ in the drag force.

If the initial velocity $V_0$ in not horizontal, the problem may be studied numerically. There is an angle range where the mentioned behavior is found. Out of that range, the velocity becomes a monotonous function and the minimum speed is the initial one (if $V_0 < V_{term}$) or the terminal one (if $V_0 > V_{term}$)

Let us assume that a projectile of mass $m$ is launched horizontally with an initial velocity $V_0$ from the top of a cliff -see Figure 1-. It is subject to the action of a gravitational field $\mathbf{g}$ and of a drag force $\mathbf{F}_d$ which is proportional to a power $n$ of the velocity. If $\theta$ is the angle between the projectile velocity and the horizontal, the equations of motion are:

$$m\frac{dV_x}{dt} = -F_d \cos\theta$$
$$m\frac{dV_y}{dt} = -F_d \sin\theta + mg \quad (1)$$

where $V_x = V\cos\theta$, $V_y = V\sin\theta$ and $F_d = kV^n$. The proportionality constant in the drag force can be expressed in terms of the terminal velocity $V_{term}$. In the equilibrium condition, $mg = kV_{term}^n$ and $k = mg/V_{term}^n$. It is convenient to use dimensionless quantities; let us introduce:

$$v_x = \frac{V_x}{V_{term}}; \quad v_y = \frac{V_y}{V_{term}}; \quad \tau = \frac{gt}{V_{term}}; \quad v = \frac{\sqrt{V_x^2 + V_y^2}}{V_{term}}$$

In this way, we may write the motion equations (1) as follow:

$$-\frac{dv_x}{d\tau} = v^{n-1} v_x$$
$$\frac{dv_y}{d\tau} = 1 - v^{n-1} v_y \qquad (2)$$

We will rewrite the previous equations to suppress the time $\tau$ from them: write $v_x$ and $v_y$ in terms of $v$ and $\theta$; then multiply the equations by $\cos\theta$ (or $\sin\theta$) and add (or subtract) them conveniently. In this way, we get a new system of equations:

$$\frac{dv}{d\tau} = \sin\theta - v^n$$
$$\frac{d\theta}{d\tau} = \frac{\cos\theta}{v} \qquad (3)$$

This new system allows us to eliminate the time $\tau$, and get a Bernoulli equation [3]:

$$\frac{dv}{d\theta} = v\tan\theta - v^{n+1}\sec\theta; \quad \theta < \pi/2 \qquad (4)$$

Notice that the above equation is not well define for $\theta = \pi/2$. Our problem is to solve eq. (4) which can be written as a linear differential equation. For doing this, a change of variable is introduced: $y = v^{-n}$. Thus, eq. (4) becomes a linear one:

$$\frac{dy}{d\theta} + ny\tan\theta = n\sec\theta \qquad (5)$$

Using standard techniques for solving linear differential equations, we finally get:

$$y = (\cos\theta)^n n \int_0^\theta \frac{d\Theta}{(\cos\Theta)^{n+1}} + y_0(\cos\theta)^n \tag{6}$$

The integral that appears above has an exact solution for any integer value of $n$ [4]. Therefore, Eq. (6) is the solution we have been looking for; it gives us the velocity evolution for the whole trajectory in a closed form. We may write the explicit solution for some values of $n$. For $n = 1$, we get:

$$v = \frac{v_0}{\cos\theta + v_0 \sin\theta} \tag{7}$$

In Fig. 2 is shown a plot of $v$ as a function of $\theta$. It is clear that the terminal velocity is reached from below regardless the value of $v_0$. The minimum velocity may be easily found, and it is:

$$v_{\min} = \left[1 + \frac{1}{v_0^2}\right]^{-1/2} < 1, \quad \forall v_0$$

This is a counter-intuitive result. One would have expected that the minimum velocity was 1, i.e. the terminal velocity.

The same result is found for other powers of $n$. For $n = 2$, the velocity is given by:

$$\frac{1}{v^2} = \sin\theta + \cos^2\theta \ln\left(\left|\frac{1 + \sin\theta}{\cos\theta}\right|\right) + \frac{1}{v_0^2} \cos^2\theta \tag{8}$$

and for $n = 3$, we get:

$$\frac{1}{v^3} = \sin\theta + 2\sin\theta \cos^2\theta + \frac{1}{v_0^3} \cos^3\theta \tag{9}$$

We have plotted eq. (9) in Fig. 3 for initial velocities closed to the terminal one. The effect is quite dramatic for those values. The projectile diminishes considerably its velocity before increasing, until the terminal velocity is reached.

Let us try to understand this phenomenon in simple terms. The projectile motion can be decomposed in the horizontal and vertical components. In both directions there is a drag force that diminishes the speeds. However, in the vertical direction the gravitational field accelerates the projectile; thus the velocity increases. At the beginning, the effect of the drag is stronger than the effect of gravity if the initial speed is large enough. But once the velocity has diminished, gravity becomes dominant and accelerates the projectile until the terminal velocity is reached. There is an interplay between two opposite causes (drag and gravity) that produces the curious behavior shown in Figs. 2 and 3.

We may wonder about what happens if the initial speed is not horizontal, i.e. if $\theta_0 \neq 0$. In this case, the ordinary differential equations displayed in (4) has been solved numerically. For that aim, a computer program was written that solves the system using a Runge-Kutta method of order fourth [5]. We have got several values of the minimum speed $v_{min}$ as a function of the initial angle $\theta_0$. See Figure 4 and 5 for $n = 2$ and $n = 3$ respectively. As can be seen from those Figures, $v_{\min} \to 1$ as $\theta_0$ increases if $v_0 > 1$, in other words, the minimum velocity is the terminal one for initial angles above a critical value ($\theta_0 \sim 50^o$). Also for $v_0 < 1$ there is a critical angle ($\theta_0 \sim 10^o$) above which the minimum velocity is the initial one. This means that the velocity is a monotonous function above those critical angles. Below the mentioned values, the velocity is non monotonous and the minimum velocity is lower than the initial one (if $v_0 < 1$) or lower than the terminal one (if $v_0 > 1$).

In summary, we have studied the trajectory of a projectile launched with horizontal velocity, and found a closed expression for the velocity as a function of the trajectory angle with the horizontal. It is surprising that the minimum velocity does not coincide with the terminal one. For any initial velocity, the projectile diminishes its speed below the terminal one, and then it accelerates and reaches the terminal velocity. This behavior is qualitatively the same for any power-law drag force. If the initial speed is not horizontal, there is critical value for the initial angle below which the same behavior is observed. Above that critical angle, the velocity becomes a monotonous function.

**Acknowledgments:** One of the authors (ENM) thanks the National Research Council of Argentina (CONICET) for support.

**Figure captions:**

**Figure 1:** This is the problem considered in the article: a projectile is launched from a cliff with an initial speed $V_0$. It moves under the action of the gravitational force $m\mathbf{g}$ and a drag force $\mathbf{F}_d$ that is proportional to a power $n$ of the speed. It is convenient to choose the angle $\theta$ as the independent variable.

**Figure 2:** The projectile velocity $v$ as a function of the angle $\theta$ (in radians) for two different initial velocities: $v_0 = 2$ (upper curve) and $v_0 = 0.5$ (lower curve). In this case, $n = 1$. Notice that the terminal velocity ($v = 1$) is different from the minimum one. The projectile trajectory is bounded in the region $0 \leq \theta \leq \pi/2$.

**Figure 3:** The projectile velocity $v$ in terms of the angle $\theta$ for initial velocities $v_0 = 1.05$ (upper curve) and $v_0 = 0.95$ (lower curve). The power in the drag force is $n = 3$. The difference between the minimum velocity and the terminal one is quite remarkable for these conditions.

**Figure 4:** This figure shows the minimum velocity $v_{\min}$ of the projectile as a function of the initial angle $\theta_0$ (in degrees) of the speed with the horizontal. The drag force in proportional to the square of the velocity ($n = 2$). The upper (lower) curve is for an initial velocity greater (lower) than the terminal one: $v_0 = 2$ ($v_0 = 0.5$). Notice that below a certain angle ($\theta_0 \sim 50°$ for $v_0 = 2$ and $\theta_0 \sim 10°$ for $v_0 = 0.5$) the minimum speed is lower than the terminal one for $v_0 = 2$, and lower than the initial one for $v_0 = 0.5$. This shows that the velocity in not a monotonous function. Above those angles, the minimum speed is either the terminal one ($v_0 = 2$) or the initial one ($v_0 = 0.5$).

**Figure 5:** This figure is the same as Figure 4 but for a drag force that goes as $F_d \sim V^3$. The minimum velocity shows the same behavior as in Figure 4.

Figure 1

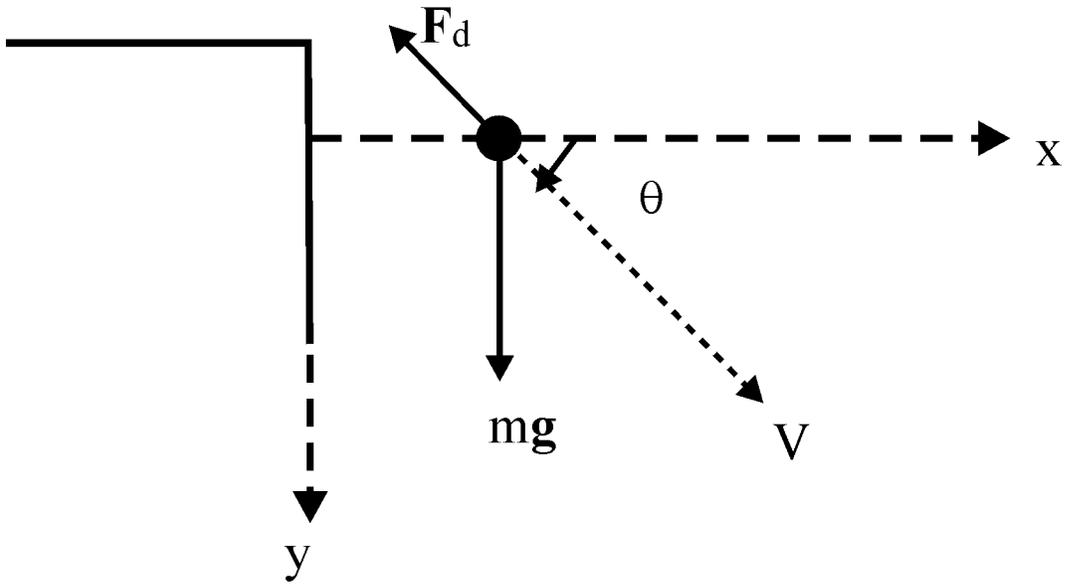

Figure 2

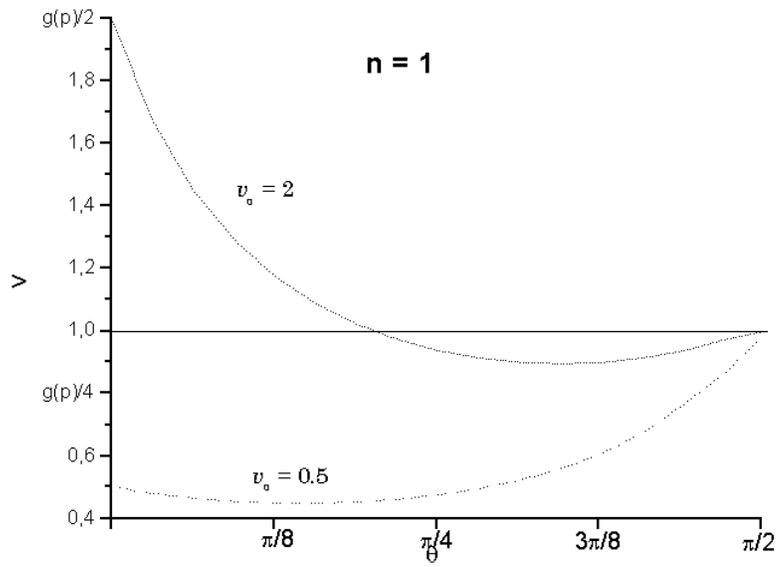

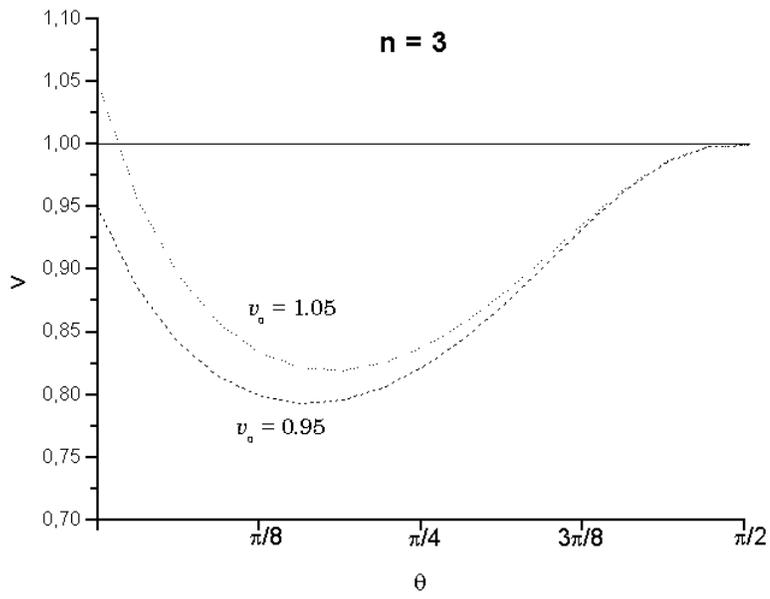

Figure 3

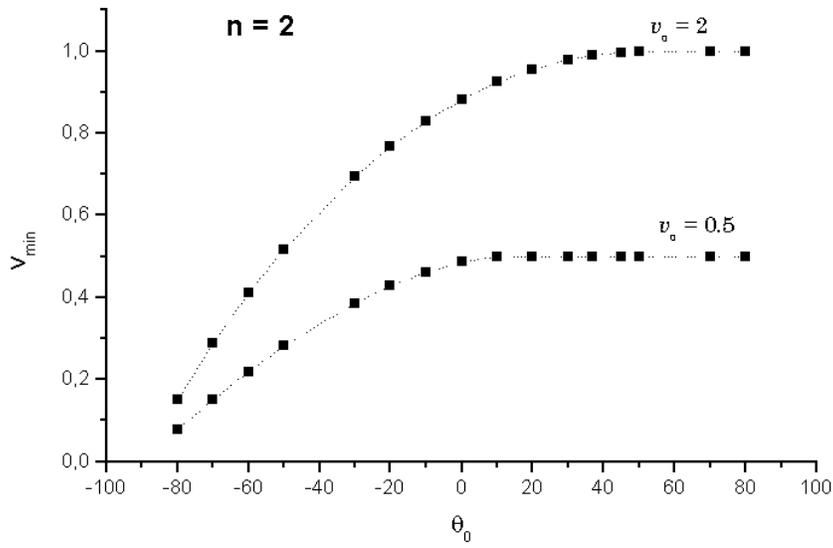

Figure 4

**Figure 5**

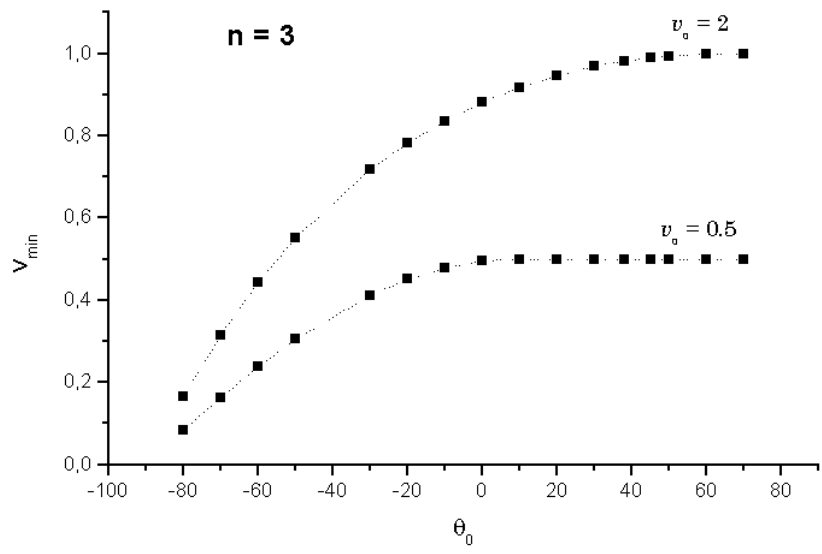